%% file: main2.tex
\documentclass{article}
\usepackage{spconf}

\input{Format.tex}
\input{Macro.tex}
\allowdisplaybreaks

\usepackage{balance}

\title{Generative Control of Time-Modulated Intelligent Reflecting Surfaces for Scalable Directional Modulation}

\name{Kailong Wang~\orcidlink{0000-0002-3415-0790}\sthanks{Kailong Wang and Sida Xiang contributed equally to this work.}, Sida Xiang~\orcidlink{0009-0005-3976-9639}\textsuperscript{*}, Athina Petropulu~\orcidlink{0000-0001-7380-7815}\thanks{The work was supported by National Science Foundation (NSF) grant ECCS-2320568 and Army Research Office (ARO) grant W911NF261A271.}}
\address{Rutgers, The State University of New Jersey, USA
    }

\begin{document}
\ninept
\maketitle

\begin{abstract}
Time-modulated intelligent reflecting surfaces (TM-IRS) achieve directional modulation by switching their elements on and off at designed instants and durations, so that an undistorted signal reaches the intended user while appearing scrambled elsewhere.
We propose a Flow Matching (FM)-based generative framework that designs these parameters for strong scrambling and reliable reception simultaneously. For orthogonal frequency-division multiplexing (OFDM) signals, we introduce a physics-informed reward that maximizes the time-modulation harmonic energy radiated in the null space of the legitimate-user response, subject to a minimum per-subcarrier signal-to-interference-plus-noise ratio (SINR), without requiring knowledge of adversary directions. Unlike prior generative flow network (GFlowNet) approaches, which assign discretized parameters sequentially, the FM model generates complete continuous configurations jointly, decoupling the generation depth from the number of elements and accelerating training. It also learns a distribution over configurations that can be resampled to vary the scrambling pattern. Numerical results show that, compared with a GFlowNet baseline, the proposed approach reduces the time to training loss near-saturation by $99.4\%$ on average for $3{\times}3$ to $12{\times}12$ surfaces, attains higher final rewards, scales to $80{\times}80$, and confines the low-symbol-error region more tightly around the legitimate user.
\end{abstract}


\input{01_Intro}
\input{02_TM-IRS}
\input{03_Flow_Matching_1}
\input{04_Experiment}

\section{Conclusion}

This paper proposed an FM framework for jointly optimizing continuous
TM-IRS timing parameters.
The method reaches the loss-based near-saturation
criterion faster than the GFlowNet baselines and supports arrays of up to
$80\times80$ elements.
We introduced a target-null sideband-energy reward with a minimum
per-subcarrier SINR gate. The revised reward promotes time modulation
while assigning zero reward to configurations that violate the reception
requirement. The desired signal phase is aligned through the fixed IRS
phase compensation.
Future work will extend the
framework to multi-user scenarios by jointly optimizing the element phase
shifts and timing parameters, increasing the number of optimization variables
from $2M$ to $3M$.


\clearpage
\balance
\bibliographystyle{IEEEbib}
\bibliography{references,npj}

\end{document}

%% file: Macro.tex
\DeclarePairedDelimiter\parens{\lparen}{\rparen}
\DeclarePairedDelimiter\bracks{\lbrack}{\rbrack}
\DeclarePairedDelimiter\braces{\lbrace}{\rbrace}
\DeclarePairedDelimiter\abs{\lvert}{\rvert}
\DeclarePairedDelimiter\norm{\lVert}{\rVert}

\newcommand*{\trn}{\!^{\mathsf{T}}}

\newcommand*{\inv}{\!^{-1}}

\newcommand*{\R}{\mathbb{R}}
\newcommand*{\C}{\mathbb{C}}

\DeclareMathOperator{\Gauv}{\mathsf{N}}

\newcommand*{\diff}{\mathop{}\!\mathrm{d}}
\newcommand*{\set}[1]{\braces*{\,#1\,}}

\makeatletter
\DeclareRobustCommand\onedot{\futurelet\@let@token\@onedot}
\def\@onedot{\ifx\@let@token.\else.\null\fi\xspace}

\def\ie{\emph{i.e}\onedot}

\makeatother

\DeclareMathOperator{\sinc}{sinc}

\newcommand\redout{\bgroup\markoverwith{\textcolor{red}{\rule[.5ex]{2pt}{0.4pt}}}\ULon}

\newtcolorbox{stretchbox}[1][]{
    height fill,
    colback=white,
    #1
    }

\newtcolorbox{problem}[1]{
    breakable,
    colbacktitle=white,
    coltitle=black,
    colback=white,
    fonttitle=\large\bfseries,
    title={#1 \hfill Grade:\hspace*{0.15\paperwidth}\ }, 
}

\newtcolorbox{solution}[1]{
    breakable,
    colback=black!3,
    fonttitle=\bfseries,
    title={#1},
}

%% file: 01_Intro.tex
\section{Introduction}\label{sec:intro}

Due to the broadcast nature of wireless propagation, confidential information may be intercepted by unintended receivers. Physical-layer security (PLS)~\cite{poor2017wireless,dong2009improving} addresses this vulnerability by exploiting the physical properties of the channel and waveform, rather than relying solely on upper-layer cryptography. Directional modulation (DM)~\cite{daly2009directional} is a PLS technique that preserves the desired signal constellation in authorized directions while intentionally scrambling the received symbols elsewhere. This capability is valuable in integrated sensing and communication (ISAC) systems~\cite{Liu2020isac, Zhang2021jcs}, where the information-bearing waveform is also radiated over a wide angular region to illuminate potential targets.
Compared with other PLS approaches, such as secrecy-rate maximization~\cite{lv2015secrecy, gong2016millimeter} or artificial-noise injection~\cite{zhang2019AN, wang2017AN}, DM can offer comparable secrecy in a more cost-effective and energy-efficient manner~\cite{su2022secure}.
Among the many DM approaches, DM enabled by a time-modulated array (TMA) driven by orthogonal frequency-division multiplexing (OFDM) signals has proven to be a hardware-efficient solution for securing wireless systems, while also enabling high data rates without the need for multiple radio-frequency (RF) chains or channel state information (CSI)~\cite{tvt2019time, tao2024tmasecurity, tao2025twc}.
Using single-pole single-throw (SPST) switches, the TMA periodically connects and disconnects its antennas from a single RF chain, generating controllable harmonics whose period is aligned with the OFDM symbol duration. These harmonics spread the symbol originally placed on each subcarrier across all subcarriers.
As a result, each subcarrier of the transmitted OFDM signal carries a weighted mixture of all original symbols, with mixing coefficients determined by the TMA parameters, namely, the switch-on instants and the corresponding on-durations. This mixing effectively scrambles the transmitted symbols. In the absence of noise, the scrambling toward an intended direction can be eliminated by a low-complexity rule-based design~\cite{tvt2019time}.

In~\cite{xu2024tmirs}, an intelligent reflecting surface (IRS)~\cite{wu2020irs} operated in a time-modulated fashion (TM-IRS) assisted an OFDM transmitter to achieve three-dimensional DM~\cite{tao2024tmasecurity, tao2025isac, tao2025secure}, with the IRS aperture compensating for the power lost while elements are off.
Its closed-form rules are easily implemented but ignore noise and guarantee undistorted reception in only one communication-user (CU) direction.
%
%
Accommodating noise and multiple CUs does not admit a closed-form solution, which motivates learning-based design. The generative artificial intelligence (GenAI) approach of~\cite{tao2025isac, tao2025secure} uses generative flow networks (GFlowNets)~\cite{bengio2023gflownet} to learn a distribution over high-performing configurations, from which distinct configurations can be sampled to vary the scrambling pattern over time.
The switching parameters are discretized and assigned sequentially, one per action,
and each configuration is scored by a reward combining the beam-pattern gain with either the legitimate-user signal-to-interference-plus-noise ratio (SINR)~\cite{tao2025secure} or the secrecy rate~\cite{tao2025isac}.
While this avoids exhaustive search over the combinatorial configuration space, the trajectory length grows with the number of control variables, a sequential dependency that parallel sampling cannot remove, and discretization restricts the switching parameters to a finite grid.


In this work, we propose a \emph{Flow Matching}~\cite{lipman2023flow} (FM)-based generative control framework for the continuous TM-IRS design problem.
Despite the shared terminology, the ``flow'' in FM is fundamentally different from that in GFlowNets.
Whereas GFlowNets define probability flows over trajectories on a directed acyclic graph, FM learns a continuous transport process between probability distributions.
Starting from random Gaussian samples, a neural vector field transports samples toward a distribution of high-reward TM-IRS configurations.
We exploit this formulation to jointly generate the complete continuous control vector, rather than constructing it from element-wise discrete decisions as in prior GFlowNet-based approaches.
All switching variables are updated simultaneously, thereby decoupling the generation depth from the number of elements.
This makes the framework scalable to large surfaces and also enables efficient online inference, since a complete configuration is produced through a short sequence of joint neural-network updates rather than a long combinatorial trajectory.
Different Gaussian initializations allow the generative controller to produce multiple distinct high-performing configurations, which is desirable for varying the spatial scrambling pattern over time.
To steer this generative process, we introduce a physics-informed reward for an OFDM transmit signal.
The reward maximizes the time-modulation harmonic energy radiated in the null space of the legitimate-user response, subject to a minimum per-subcarrier signal-to-interference-plus-noise ratio at that user.
In this way, no information about the directions to be protected is required.

%% file: 02_TM-IRS.tex
\section{System Model and Problem Formulation}\label{sec:system}

\begin{figure}
\centering
\includegraphics[width=\columnwidth]{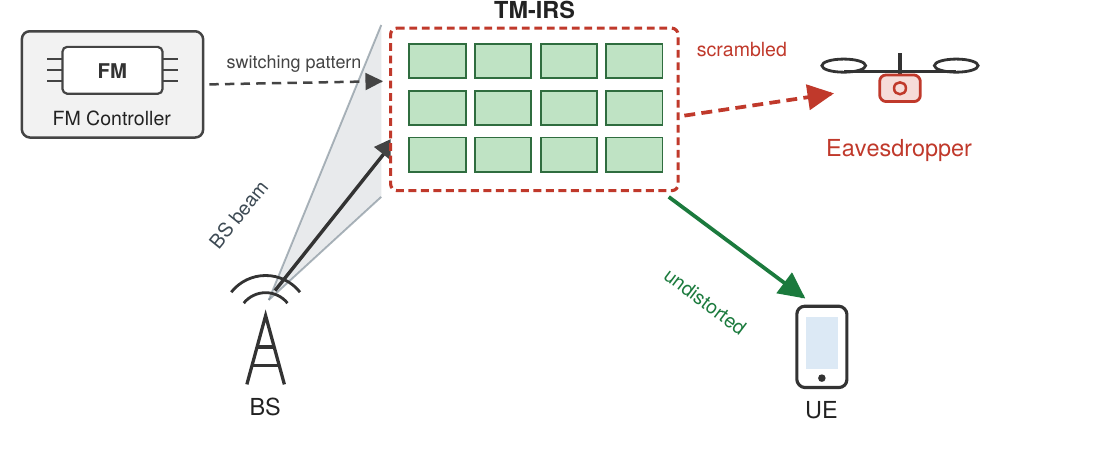}
\caption{TM-IRS-assisted system with the IRS adjacent to the BS for collaborative beamforming and configured by an FM controller.
}
\label{fig1}
\end{figure}

We consider an $N$-element uniform linear array (ULA) transmitting through an $M_x\times M_z$ TM-IRS, where $M=M_xM_z$ is the number of reflecting elements (see \cref{fig1}).
With the ULA beamformed toward the IRS, the incident OFDM waveform is $e\parens{t}=\sqrt{\frac{N}{K}}\sum_{k=0}^{K-1}d_k e^{j2\pi\parens{f_c+kf_s}t}$, $0\leq t<T_s$, where $K$ is the number of subcarriers, $d_k$ is the data symbol on subcarrier $k$, $f_c$ is the carrier frequency, $f_s$ is the subcarrier spacing, and $T_s$ is the symbol duration.

With half-wavelength spacing between adjacent IRS elements along both array dimensions, the far-field response of IRS element $(m,n)$ toward $(\theta,\phi)$ is $a_{mn}\parens{\theta,\phi} =\exp\braces{-j\pi\parens{m\sin\theta\cos\phi+n\sin\theta\sin\phi}}$, where $m=0,\ldots,M_x-1$ and $n=0,\ldots,M_z-1$.
Let $(\theta_T,\phi_T)$ denote the transmitter direction relative to the IRS.
Each IRS element applies a unit-modulus phase shift $c_{mn}$ and is periodically switched with normalized switch-on instant $\tau^o_{mn}\in[0,1)$ and activation duration $\Delta\tau_{mn}\in[0,1)$.
After propagation to the IRS and reflection toward $(\theta,\phi)$, the received waveform is~\cite{tao2025secure}
\begin{multline}
    y\parens{t,\theta,\phi}
    =
    \sqrt{\frac{N}{K}}
    \sum_{k=0}^{K-1}d_k e^{j2\pi\parens{f_c+kf_s}t} \\
    \times
    \sum_{l=-\infty}^{\infty}
    V_l\parens{\theta,\phi}e^{j2\pi l f_s t}
    +z\parens{t},
    \label{eq:received_waveform}
\end{multline}
where $z\parens{t}$ denotes receiver noise and
\begin{multline}
    V_l\parens{\theta,\phi}
    =
    \sum_{m=0}^{M_x-1}\sum_{n=0}^{M_z-1} a_{mn}\parens{\theta_T,\phi_T}c_{mn}a_{mn}\parens{\theta,\phi}\\
    \times
    \Delta\tau_{mn}\sinc\parens{l\pi\Delta\tau_{mn}} e^{-jl\pi\parens{2\tau^o_{mn}+\Delta\tau_{mn}}},
    \label{eq:harmonic_coefficient}
\end{multline}
is the effective coefficient of the $l$-th time-modulation harmonic.
Since the harmonic spacing equals $f_s$, these harmonics mix the OFDM subcarriers.
After demodulation, the $l$-th time-modulation harmonic generated from subcarrier $k$ contributes to subcarrier $i=k+l$, \ie, $l=i-k$.
Specifically, the received symbol on subcarrier $i$ is
\begin{equation}
    y_i\parens{\theta,\phi}
    =
    \sqrt{\frac{N}{K}}
    \sum_{k=0}^{K-1}
    d_k V_{i-k}\parens{\theta,\phi}+z_i ,
    \label{eq:demodulated_symbol}
\end{equation}
where $V_0$ scales the desired symbol, and $V_{i-k}$, $k\neq i$ produce inter-subcarrier interference.

For the single-user setting, let $(\theta_c,\phi_c)$ denote the legitimate-user direction.
We choose~\cite{xu2024tmirs} $c_{mn} = \bracks*{ a_{mn}\parens{\theta_T,\phi_T} a_{mn}\parens{\theta_c,\phi_c} }\inv$, which cancels the spatial phase at the legitimate user.
Therefore, only the $2M$ timing variables $\set{\tau^o_{mn},\Delta\tau_{mn}}_{m,n}$ are optimized.
Writing $V_l=V_l\parens{\theta_c,\phi_c}$ for brevity, we obtain
\begin{equation}
    V_l
    =
    \sum_{m=0}^{M_x-1}\sum_{n=0}^{M_z-1}
    \Delta\tau_{mn}\sinc\parens{l\pi\Delta\tau_{mn}}
    e^{-jl\pi\parens{2\tau^o_{mn}+\Delta\tau_{mn}}}.
    \label{eq:user_harmonic}
\end{equation}
Assuming unit-power symbols and noise variance $\sigma^2$, the per-subcarrier SINR, with $\eta=N/K$, is
\begin{equation}
    \mathrm{SINR}_i
    =
    \frac{\eta \abs{V_0}^2}
    {\eta\displaystyle\sum_{\substack{k=0,k\neq i}}^{K-1}
    \abs{V_{i-k}}^2+\sigma^2},
    \quad i=0,\ldots,K-1.
    \label{eq:sinr}
\end{equation}
The achievable sum rate across all subcarriers, given by
\begin{equation}\label{eq:sumrate}
    C_{\mathrm{achievable}} =\sum_{i=0}^{K-1}\log_2\parens{1+\mathrm{SINR}_i},
\end{equation}
served as the reward for GFlowNet training~\cite{tao2025secure}.
However, under continuous timing control,  this objective admits an \emph{always-on degeneration}.
Since $0\leq\Delta\tau_{mn}<1$, \cref{eq:sinr} satisfies $\mathrm{SINR}_i<{\eta M^2}/{\sigma^2}$.
As $\Delta\tau_{mn}\rightarrow1$ for all $(m,n)$, $V_0\rightarrow M$, while $\sinc\parens{l\pi\Delta\tau_{mn}}\rightarrow0$ for every $l\neq0$.
Consequently, the legitimate-user SINRs approach the SINR bound  while the nonzero harmonics responsible for scrambling vanish, and the TM-IRS approaches an unswitched static IRS.
Thus, the sum rate approaches its upper bound while the scrambling induced by time modulation vanishes.
In the discretized design of \cite{tao2025secure}, a fixed $Q$-point discretization restricts the maximum duration to $(Q-1)/Q$, maintaining a gap of $1/Q$ from this endpoint.
Although the always-on endpoint is excluded from the interval $[0,1)$, continuous timing control could allow the activation durations to approach it closely.
To address this issue, \cref{sec:flow_control} develops a physics-informed reward that promotes nonzero-harmonic energy and assigns zero reward to configurations that do not meet the reception requirement.

%% file: 03_Flow_Matching_1.tex
\section{Flow-Matching-Based TM-IRS Control}\label{sec:flow_control}

Each TM-IRS element is controlled by two continuous timing parameters: the normalized switch-on instant $\tau^o_{mn}$ and switch-on duration $\Delta\tau_{mn}$.
For an IRS with $M$ elements, these parameters form a configuration vector $\bm{x}\in[0,1)^{2M}$ by stacking the pairs $\parens{\tau^o_{mn},\Delta\tau_{mn}}$ in a fixed element order.
We use an FM-based model~\cite{lipman2023flow} to learn a vector field that transforms Gaussian random inputs into terminal states, which are then mapped to TM-IRS timing configurations.
This generation procedure defines a stochastic policy from which candidate configurations can be sampled.

The model learns a vector field $v_{\bm{\theta}}$ through reward-guided policy optimization, as described in \cref{sec:rewardandopt}.
Starting from a Gaussian random vector $\bm{z}_0\sim\Gauv\parens{\bm{0},\bm{I}_{\rm 2M}}$, the intermediate state $\bm{z}_t\in\R^{2M}$ changes according to ${\diff\bm{z}_t}/{\diff t} =v_{\bm{\theta}}\parens{\bm{z}_t,t}$, $t\in[0,1]$, where $\bm{\theta}$ denotes the network parameters.
Given $\bm{z}_t$ and $t$, the network outputs the rate of change of all $2M$ state components.
We generate samples using $L$ Euler steps with step size $h=1/L$: $\bm{z}_{t_{k+1}} =\bm{z}_{t_k}+h\,v_{\bm{\theta}}\parens{\bm{z}_{t_k},t_k}$, $t_k=kh$, for $k=0,\ldots,L-1$.
The terminal state $\bm{z}_1$ is then mapped to the timing configuration $\bm{x}$.
Here, $t$ describes progress from initialization at $t=0$ to completion at $t=1$.

The internal state $\bm{z}_t$ takes values in $\R^{2M}$, whereas the physical timing configuration $\bm{x}$ must lie in $[0,1)^{2M}$.
We therefore map the terminal state $\bm{z}_1$ to $\bm{x}$ using a sigmoid function applied to each component: $\bm{x} =\operatorname{sigmoid}\parens{\bm{z}_1} ={1}/{(1+\exp\parens*{-\bm{z}_1})}$.
At each Euler step, the FM model jointly updates all $2M$ state components corresponding to the TM-IRS timing parameters.
For a fixed number of Euler steps, increasing the IRS size enlarges the state vector without introducing additional element-wise sequential decisions.




\subsection{Physics-Informed Reward}\label{sec:rewardandopt}

Directly maximizing the legitimate-user sum rate does not sufficiently promote the nonzero harmonics required for directional scrambling, resulting in the always-on degeneration as explained in \cref{sec:system}.
We therefore construct the reward as follows.

For harmonic order $l\neq0$, define the element-wise harmonic vector $\bm{b}_l = \bracks*{ b_{00}\parens{l},\ldots,b_{M_x-1,M_z-1}\parens{l} }\trn$, where $b_{mn}\parens{l}=c_{mn}\Delta\tau_{mn}\sinc\parens{l\pi\Delta\tau_{mn}}e^{-jl\pi\parens{2\tau^o_{mn}+\Delta\tau_{mn}}}$.
For the legitimate-user direction, define $\bm{s}=\bracks{s_{00},\ldots,s_{M_x-1,M_z-1}}\trn\in\C^M$, where $s_{mn}=a_{mn}\parens{\theta_T,\phi_T}a_{mn}\parens{\theta_c,\phi_c}$.
This is the effective steering vector for the transmitter--IRS--legitimate-user path.
Each entry combines the incident and reflected spatial phase responses of the corresponding IRS element, as defined in \cref{sec:system}.

From \cref{eq:harmonic_coefficient} and the definitions of $\bm{b}_l$ and $\bm{s}$, we have $V_l=\bm{s}\trn\bm{b}_l$.
To obtain the harmonic component in the null space of the legitimate-user response, we use the orthogonal projector $\bm{P}_{\perp} = \bm{I}_{\rm M} - \frac{\bm{s}^*\bm{s}\trn}{\norm{\bm{s}}_2^2}$, which satisfies $\bm{s}\trn\bm{P}_{\perp}=\bm{0}\trn$.
Since the entries of $\bm{s}$ have unit magnitude, the projected harmonic energy can be written as $\norm{\bm{P}_{\perp}\bm{b}_l}_2^2 =\norm{\bm{b}_l}_2^2-\frac{\abs{V_l}^2}{M}$.
This component does not contribute to the received harmonic at the legitimate user.
We then define the sideband objective as $E_{\perp} = \sum_{l\in\mathcal{L}_{\neq0}} \norm{\bm{P}_{\perp}\bm{b}_l}_2^2$, where $\mathcal{L}_{\neq0}$ is the set of nonzero harmonic orders included in the objective.
This objective promotes harmonic energy in the legitimate-user null space and provides a measure for designing directional scrambling.
To incorporate the minimum SINR requirement on every subcarrier, we define the reward as
\begin{equation}
    R\parens{\bm{x}}
    =
    E_{\perp}
    \,\mathcal{H}\!\parens*{
    \min_{0\leq i<K}\mathrm{SINR}_i-\Gamma_{\min}
    },
    \label{eq:reward}
\end{equation}
where $\mathcal{H}\parens{u}=1$ for $u\geq0$ and zero otherwise, and $\Gamma_{\min}>0$ is the required SINR threshold.
Configurations that violate the SINR requirement receive a zero reward, encouraging the policy to generate feasible configurations, even though individual samples may still violate the requirement.

At each training iteration, the FM policy generates $G$ groups of candidate configurations from Gaussian noise, with each group containing $S$ samples.
Let $\bm{z}_{gs}\in\R^{2M}$ denote the flow endpoint before mapping for the $s$-th sample in the $g$-th group, which is mapped to the physical timing configuration $\bm{x}_{gs}$.
The corresponding reward is evaluated using \cref{eq:reward} and denoted by $R_{gs}=R\parens{\bm{x}_{gs}}$, which updates the FM policy using Flow Policy Optimization (FPO)~\cite{mcallister2026fpo}.

%% file: 04_Experiment.tex
\section{Experiments and results}
\label{sec:results}

\begin{figure*}
    \centering
    \subfloat[FM with sum-rate reward. \label{fig:fm_sumrate_fm}]{
        \includegraphics[width=0.33\linewidth]{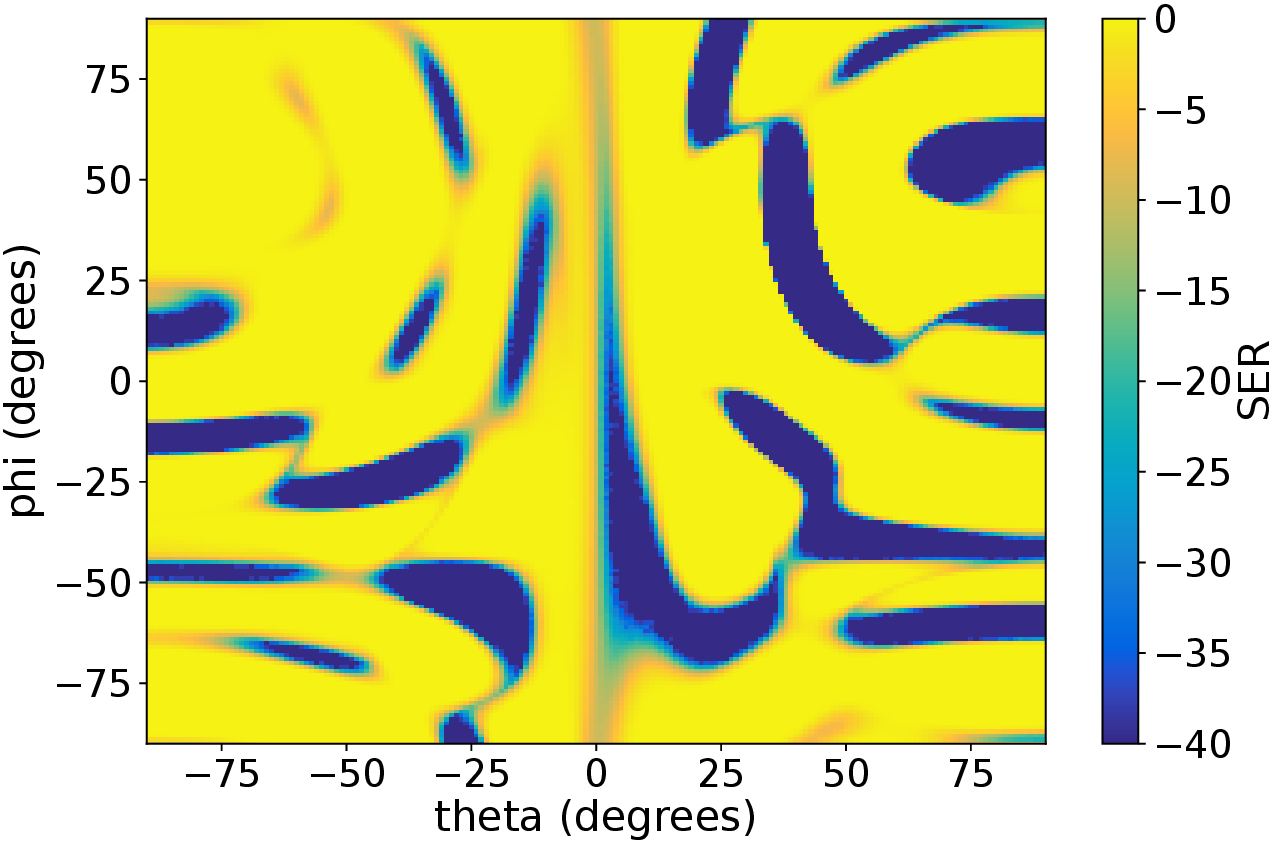}
        }%
    \subfloat[GFN with sum-rate reward. \label{fig:gfn-sumrate_fm}]{
        \includegraphics[width=0.33\linewidth]{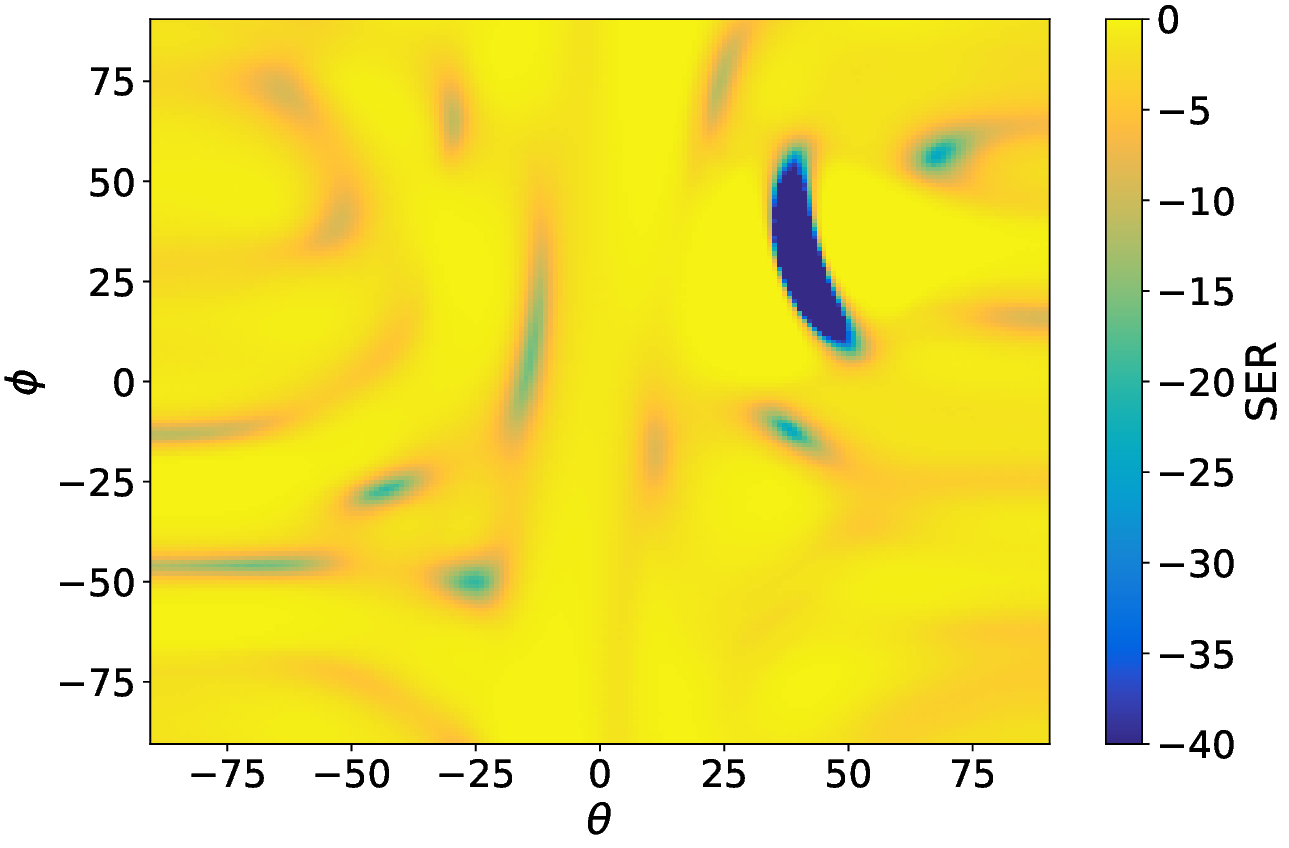}
        }%
    \subfloat[FM with the proposed reward. \label{fig:proposed_fm}]{
        \includegraphics[width=0.33\linewidth]{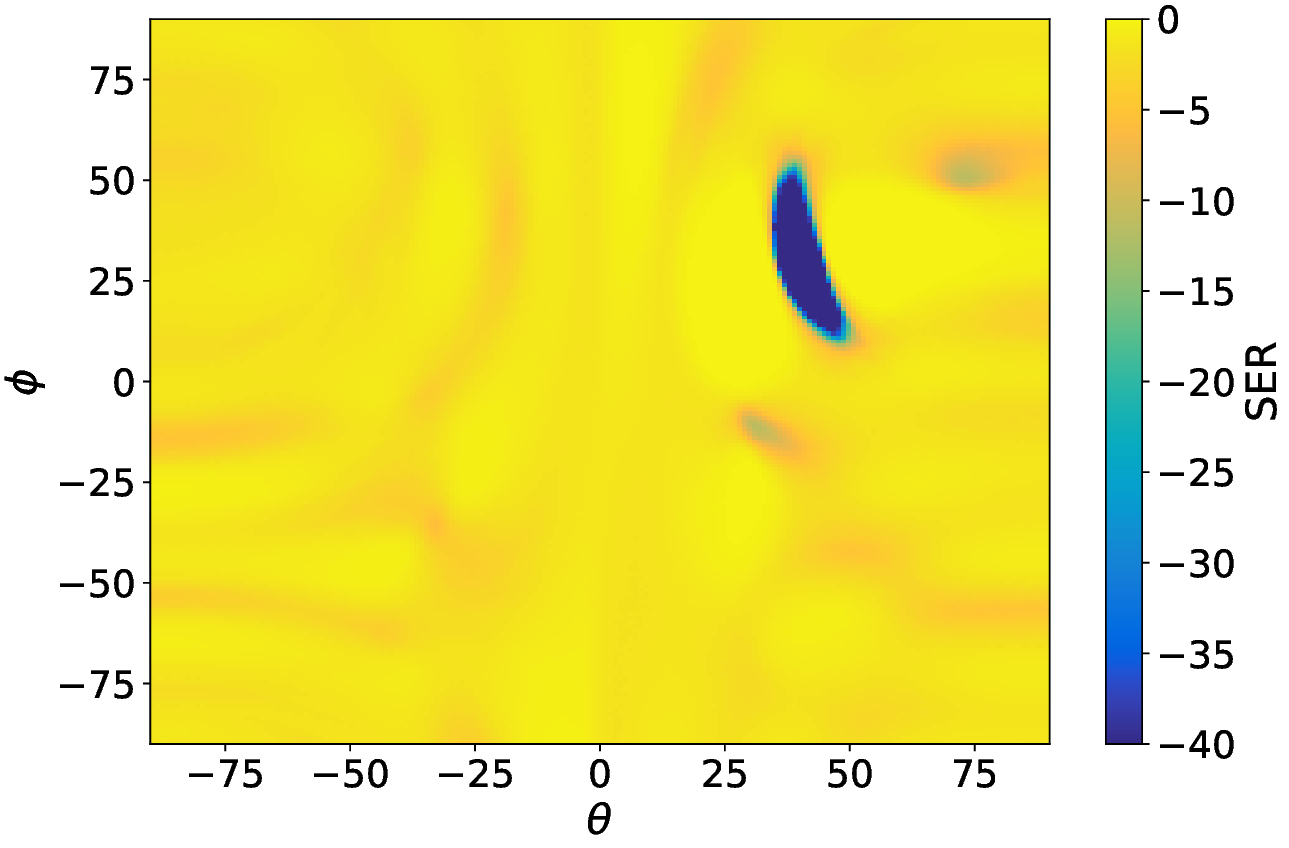}
        }%
    \caption{SER maps for a $6\times6$ TM-IRS using GFN and FM.}
    \label{fig:reward_comparison}
\end{figure*}

\begin{figure*}
    \centering
    \subfloat[Final reward across IRS sizes. \label{bb}]{
    \includegraphics[width=0.33\linewidth]{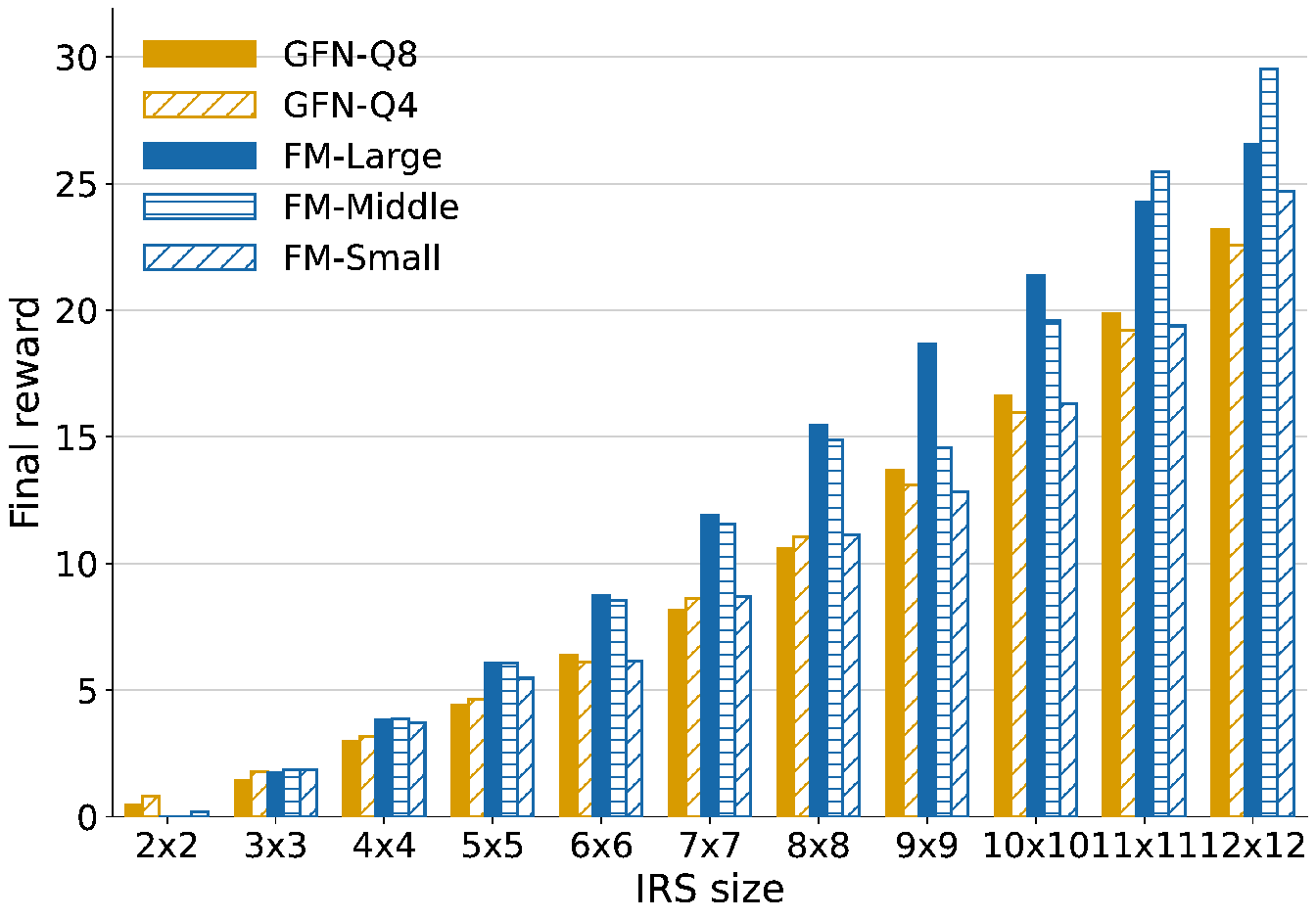}
    }%
    \subfloat[Time to near-saturation. \label{aa}]{
    \includegraphics[width=0.33\linewidth]{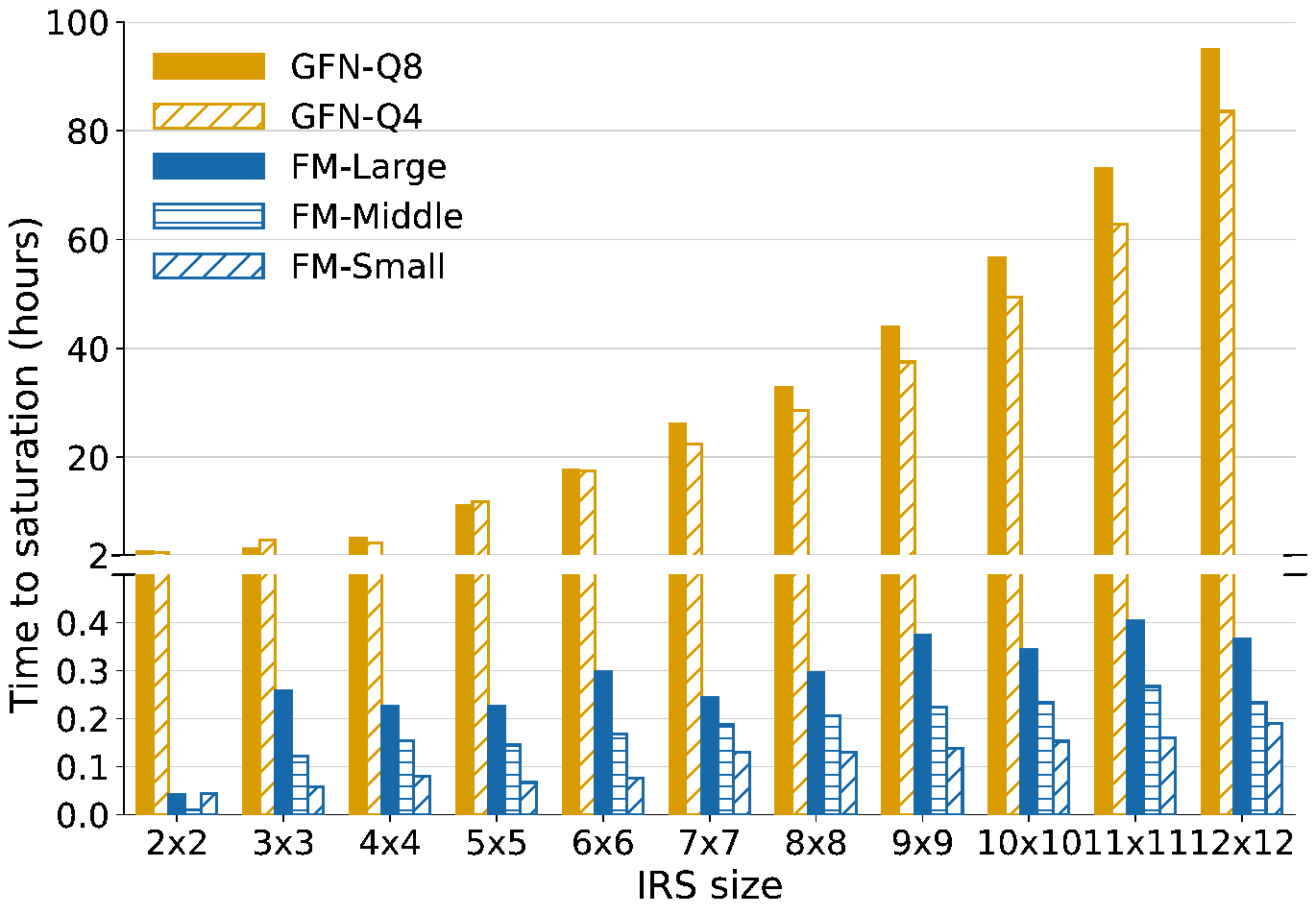}
    }%
    \subfloat[Final reward and SINR pass rates across IRS sizes. \label{cc}]{
    \includegraphics[width=0.33\linewidth]{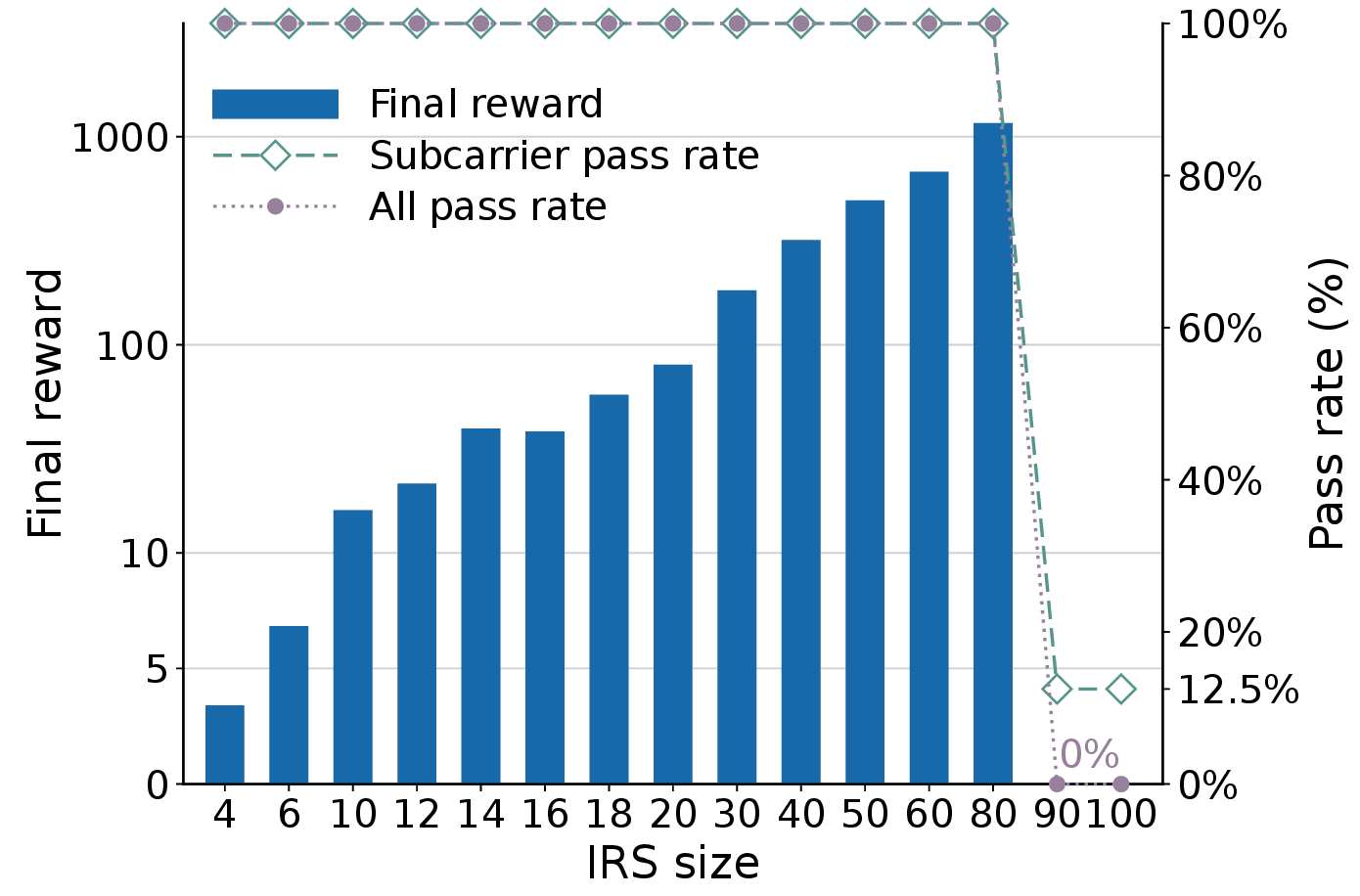}
    }%
    \caption{Training efficiency and performance across TM-IRS sizes. }
    \label{fig:training_time_comparison}
\end{figure*}

We consider a single-user OFDM system with $N=8$ transmit antennas and $K=16$ subcarriers, with the nonzero harmonic orders $\mathcal L_{\neq0}=\set{\pm 15,\ldots,\pm1}$ included in the proposed reward.
Quadrature phase-shift keying (QPSK) signaling is used with a signal-to-noise ratio (SNR) of $0$~dB and unit path gain.
The transmitter and legitimate user are located at $(\theta_T,\phi_T)=(15^\circ,10^\circ)$ and $(\theta_c,\phi_c)=(40^\circ,30^\circ)$, respectively.
For an $M_x\times M_z$ TM-IRS with $M=M_xM_z$ elements, the phase coefficients are fixed as $c_{mn}=\bracks*{a_{mn}\parens{\theta_T,\phi_T}a_{mn}\parens{\theta_c,\phi_c}}\inv$.
The proposed reward imposes a per-subcarrier SINR threshold of $14$~dB, corresponding to $\Gamma_{\min}=10^{14/10}$.
We evaluate the symbol error rate (SER) using 1,024 OFDM symbols and LMMSE detection at observation directions $(\theta,\phi)$ on an angular grid with a $1^\circ$ step in each coordinate.
The results are displayed as angular maps of SER in dB, $10\log_{10}(\mathrm{SER})$, where darker colors indicate easier detection and lighter colors indicate more difficult detection.

The FM policy jointly optimizes all $2M$ timing variables $\set{\tau^o_{mn},\Delta\tau_{mn}}_{m,n}$, using $L=10$ to generate each configuration.
The velocity network has three hidden layers, with 128, 256, and 512 neurons per layer for FM-Small, FM-Middle, and FM-Large, respectively.
We train the policy using FPO with the Adam optimizer and a learning rate of $3\times10^{-4}$.
At each iteration, we sample $G=64$ groups under the same fixed legitimate-user direction, with $S=8$ configurations per group, yielding 512 configurations in total.
Each batch is used for four update epochs, with minibatch size 128 and an FPO clipping parameter of $0.05$.
All training runs are performed on the Rutgers Amarel high-performance computing (HPC) cluster.

\cref{fig:reward_comparison} illustrates the \emph{always-on degeneration} of the sum-rate reward in \cref{eq:sumrate} and its resolution by the proposed reward on a $6\times6$ TM-IRS.
Under the sum-rate reward, FM drives the switch-on durations toward unity: $32$ of the $36$ elements ($88.9\%$) have $\Delta\tau_{mn}>0.99$.
The nonzero harmonics responsible for scrambling are therefore suppressed, leaving the symbols easy to detect in unintended directions (see \cref{fig:fm_sumrate_fm}).
The proposed reward avoids this degeneration, yielding high SER in unintended directions while maintaining low SER at the legitimate user (see \cref{fig:proposed_fm}).
%
%
%
%
{
\cref{fig:gfn-sumrate_fm} shows the corresponding SER map for the GFlowNet method of~\cite{tao2025secure}, which optimizes discretized parameters under the sum-rate reward.
Although its discretization caps the switch-on durations at $(Q-1)/Q$ and thus avoids the always-on degeneration,
the GFlowNet configuration still exhibits more low-SER regions in unintended directions than the FM configuration in \cref{fig:proposed_fm}, indicating weaker directional scrambling.
}



{We next compare the final rewards of the FM variants and the GFlowNet baselines, and examine how FM network width affects performance.}
\cref{bb} compares the final validation rewards of FM-Small, FM-Middle, and FM-Large against two GFlowNet baselines across IRS sizes from $2\times2$ to $12\times12$.
The baselines, GFN-Q4 and GFN-Q8, use four and eight discrete levels per optimized parameter, respectively, and both have three hidden layers with 256 neurons per layer.
All five methods use the proposed reward, and the reported rewards are averaged over three independent training seeds.
For IRS sizes from \(3\times3\) onward, FM-Middle and FM-Large generally achieve higher rewards than the GFlowNet baselines, whereas FM-Small shows less consistent gains. This suggests that wider FM networks can improve configuration quality under the proposed reward. However, the improvement is not monotonic in network width: FM-Middle outperforms FM-Large at \(11\times11\) and \(12\times12\). At \(2\times2\), the FM models mostly failed to find feasible configurations.


\cref{aa} compares the time to near-saturation from the same runs, averaged over three training seeds.
The reward in \cref{eq:reward} measures configuration quality and provides feedback for training.
The training loss is the method-specific optimization objective: the trajectory-balance loss for GFlowNet~\cite{tao2025secure} and the FPO loss for FM~\cite{mcallister2026fpo}.
Our near-saturation criterion measures the stabilization of each method's training loss; it does not imply equal final rewards across methods.
We define near-saturation as the earliest point from which at least $90\%$ of the remaining smoothed training loss values lie within $[m-d,m+d]$.
Here, $m$ is the median of the final $10\%$ of the smoothed training loss
values, and $d$ is the $95$th percentile of the absolute deviations
of those final values from $m$.
We measure time to near-saturation as the cumulative wall-clock
training time up to this point.
The $2\times2$ results are included in the figure, but the FM models failed to find feasible configurations for most training seeds at this size, so we omit this case when calculating the average time reduction.
Across the ten IRS sizes from $3\times3$ to $12\times12$, FM-Small, FM-Middle, and FM-Large reduce the time to near-saturation by an average of $99.4\%$, $98.8\%$, and $98.0\%$, respectively, relative to GFN-Q8, and by $99.4\%$, $98.8\%$, and $98.1\%$, respectively, relative to GFN-Q4.
All three FM variants therefore reach the prescribed criterion substantially sooner than GFN-Q8, with FM-Small achieving the largest average reduction.


Finally, we evaluate whether the trained FM-Small models satisfy the minimum per-subcarrier SINR requirement at the legitimate user.
\cref{cc} reports the mean reward and SINR pass rates over $500$ generated configurations.
The subcarrier pass rate is the fraction of the $8{,}000$ subcarrier observations meeting the $14$~dB threshold, and the all-pass rate is the fraction of configurations meeting it on all $16$ subcarriers.
From $4\times4$ to $80\times80$, both pass rates are $100\%$ and the mean reward generally increases with array size, as does the reward variance (\cref{tab:variance}).
At $90\times90$ and $100\times100$, the reward and its variance are zero throughout training and evaluation, with subcarrier and all-pass rates of $12.5\%$ and $0\%$, respectively.
This is a limitation of the hard SINR gate in \cref{eq:reward}, rather than of the FM framework itself: no feasible configuration is sampled during early training, so all rewards within each group are zero, all advantage weights  vanish, and the policy receives no learning signal.
Since the SINR bound $\eta M^2/\sigma^2$ grows with $M$, feasible configurations are expected to exist at these sizes. Replacing the hard gate with a graded penalty on the SINR shortfall, or gradually raising $\Gamma_{\min}$ during training, would restore a within-group ranking signal and is left for future work.


\input{variance}

%% file: variance.tex
\begin{table}[t]
    \centering
    \caption{Reward variance across TM-IRS sizes.}
    \label{tab:variance}
    \begin{adjustbox}{max width=\columnwidth}
    \begin{tabular}{@{}cc@{\qquad}cc@{\qquad}cc@{}}
        \hline
        IRS size & Variance & IRS size & Variance & IRS size & Variance \\
        \hline
        $4\times4$  & $6.8\times 10^{-7}$ & $16\times16$ & 0.092 & $50\times50$  & 1.7 \\
        $6\times6$  & 0.0066              & $18\times18$ & 0.16  & $60\times60$  & 2.4 \\
        $10\times10$ & 0.031               & $20\times20$ & 0.22  & $80\times80$  & 4.4 \\
        $12\times12$ & 0.036               & $30\times30$ & 0.52  & $90\times90$  & 0   \\
        $14\times14$ & 0.091               & $40\times40$ & 1.0   & $100\times100$ & 0   \\
        \hline
    \end{tabular}
    \end{adjustbox}
\end{table}